\documentclass[aps,pre,twocolumn,showpacs,superscriptaddress,groupedaddress]{revtex4-1}
\pdfoutput=1
\usepackage{hyperref}
\usepackage{epsfig}
\usepackage{amsmath}
\usepackage{graphicx}
\usepackage{wrapfig}
\usepackage{dcolumn}
\usepackage{bm}
\usepackage{amssymb}
\usepackage{titlesec}
\usepackage{mathtools}
\usepackage{relsize}
\usepackage{cleveref}
\usepackage[usenames,dvipsnames]{color}
\usepackage{ulem}

\def\k{\textcolor{ForestGreen}}

\begin{document}
\title{Hybrid dynamics in delay-coupled swarms with ``mothership" networks}
\author{Jason Hindes}
\author{Klementyna Szwaykowska}
\author{Ira B. Schwartz}
\affiliation{U.S. Naval Research Laboratory, Code 6792, Plasma Physics Division, Nonlinear Dynamical Systems Section, Washington, DC 20375}
\begin{abstract}

Swarming behavior continues to be a subject of immense interest because of its centrality
in many naturally occurring systems in physics and biology, as well as its importance in applications such as robotics.
%Moreover, the
%development of autonomous mobile agents that can mimic the behavior of swarms
%and can be engineered to perform complex tasks without constant intervention
%is a very active field of research.
Here we examine the effects on swarm
pattern formation from  delayed communication and topological heterogeneity, and in particular, the inclusion of a relatively small number of highly connected nodes,
or ``motherships", in a swarm's communication network. We find generalized forms of basic patterns for networks with general degree distributions, and a variety of new behaviors including new parameter regions with multi-stability and hybrid motions in bimodal networks. The latter is an interesting example of how heterogeneous networks can have  dynamics that is a mix of different states in homogeneous networks, where high and low-degree nodes have simultaneously distinct behavior.
\end{abstract}
\maketitle

\section{\label{sec:Intro} INTRODUCTION}
Much attention has been given to the study of multi-agent swarms in natural and
engineered systems that can self organize and form complex spatiotemporal
patterns from very basic rules governing individual dynamics \cite{Vicsek,Marchetti,Aldana}.
This is motivated by many fascinating phenomena such as schooling
fish, flocking birds, swarming locusts, and colonizing bacteria \cite{Budrene,Polezhaev,Tunstro}. Also of
great interest is the application of principles underlying such behaviors to
the design of networks of autonomous robots and mobile sensors, with the aim of producing scaleable
and robust swarms that can perform complicated tasks without constant human intervention \cite{Bandyopadhyay,Wu}.

Several recent works in swarm pattern formation have focused on time-delay effects, which can produce new patterns and bistability \cite{Romero, Romero2, Forgoston}.
Delays model the finite time required for agents to send and process information in real systems. They are ubiquitous in both natural and engineered settings, and often can be comparable to other relevant timescales. Salient examples of natural systems where delays can significantly affect the dynamical behaviors include bat flights; predator-prey population dynamics; and blood cell production \cite{Gig, Martin, Monk}. Significant delays can also occur in robotic swarms communicating over wireless networks with low bandwidth \cite{Stachura2011}, which must be taken into account when considering the stability of swarming or formation-keeping behaviors \cite{Franco2008,Viragh2014} and performance for tasks such as swarm teleoperation \cite{Liu2015}.

Most studies of multiagent robotic systems with time delay have assumed global
interactions or homogeneous topology \cite{Romero, Forgoston, Klimka}. In general, the effects of complex network structure
on swarm behavior are much less explored, particularly with time-delayed interactions, even though topology is known to strongly influence
many processes and produce interesting new dynamics \cite{Vicsek,Dorogovtsev,Vespignani1,Hindes2}. Here, we focus on delay-coupled swarms interacting through heterogeneous networks that have a relatively small fraction of highly connected nodes, or ``motherships." Such nodes can mimic the influence of leaders in social networks or the insertion of highly interacting controllers into a network of autonomous mobile robots-- intended to alter the motion to a different form.

To understand dynamic pattern formation in swarms with delay and heterogeneity, we consider a general model for $N$ interacting, self-propelled agents \cite{Erdmann}:
%\vspace{-0.4in}
\begin{align}
\label{eq:Dynamics}
&{\boldsymbol{\ddot{r}_{i}}}(t)=(1- \dot{r}_{i}^{2}){\boldsymbol{\dot{r}_{i}}} \nonumber \\
&-J\sum\limits_{j=1}^N A_{ij}({\boldsymbol{{r}_{i}}}(t),{\boldsymbol{{r}_{j}}}(t-\tau)) \nabla_{{\boldsymbol{{r}_{i}}}}U({\boldsymbol{{r}_{i}}}(t)-{\boldsymbol{{r}_{j}}}(t-\tau)),
\end{align}
where ${\boldsymbol{{r}_{i}}}$ is the position of the $i$th agent in $2$-dimensions, dots denote time derivatives, $A_{ij}$ is the connection matrix, $J$ is the coupling strength between neighbors in the network, and $\tau$ is the characteristic time delay between agent interactions \cite{Romero2,Klimka}. For simplicity, we assume that the mutual forces are spring-like:
$\nabla_{{\boldsymbol{{r}_{i}}}}U({\boldsymbol{{r}_{i}}}(t)-{\boldsymbol{{r}_{j}}}(t-\tau))={\boldsymbol{{r}_{i}}}(t)-{\boldsymbol{{r}_{j}}}(t-\tau)$, though sufficiently small repulsive terms do not alter the dynamics \cite{Romero}.

In this work, we discuss the behaviors for Eq.(\ref{eq:Dynamics}) given simple heterogeneous topology. In addition to generalizing the patterns from homogeneous networks, we show that heterogeneity can produce novel hybrid motions, where different parts of the network have different collective dynamics depending on the degree of local connectivity. The production of new states that are mixes of distinct behaviors for homogeneous networks is an interesting feature of nonlinear processes occurring on heterogeneous networks and is seen in other contexts, e.g., coupled oscillators \cite{Hindes2}. Hybrid behaviors are practically interesting in this context as well because they offer the possibility for synthetic swarms to perform multiple tasks simultaneously. Different mechanisms for swarm splitting behaviors, observed in swarming systems with no communication delay, are described e.g. in \cite{Chen2010,Zhao2011}.

\begin{figure*}[t]
\centering
\includegraphics[scale=0.22]{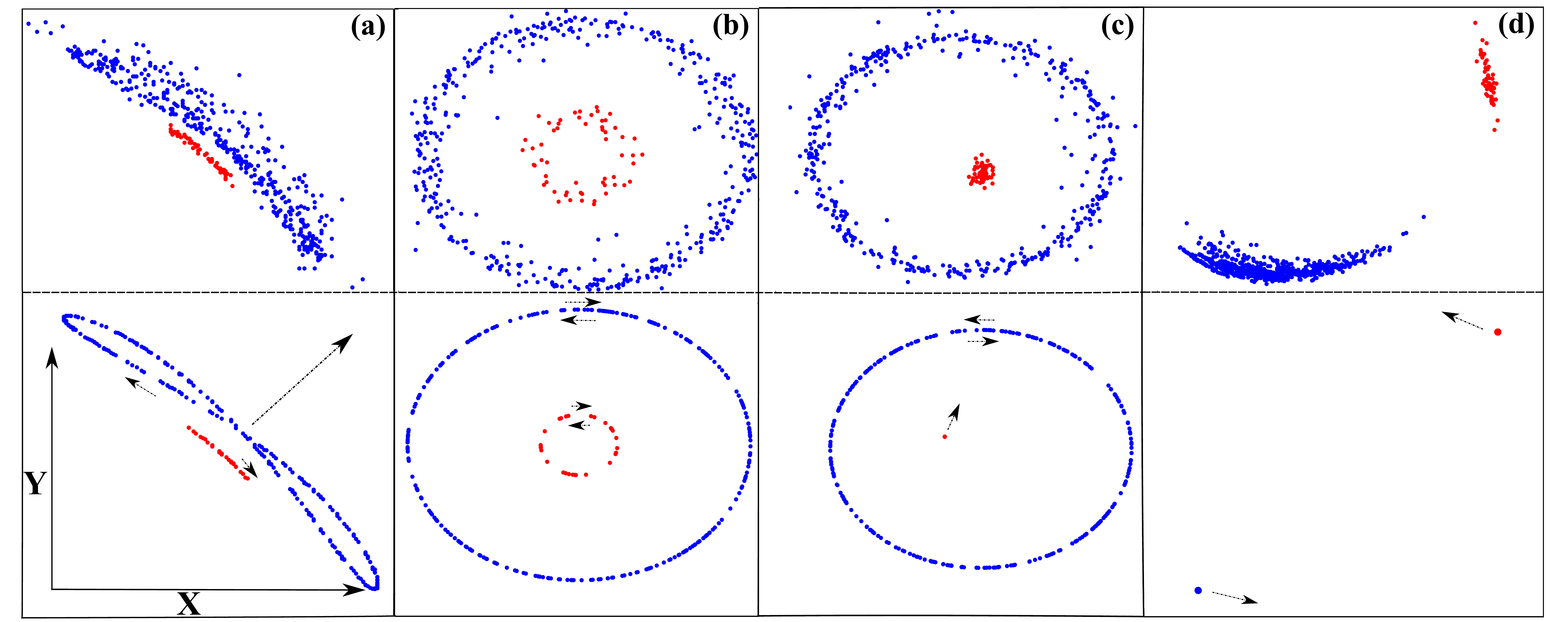}
\caption{{Patterns for swarms with time-delayed interactions and an underlying network with a small fraction of highly connected nodes, or ``motherships".
 $CM$ and annealed network patterns are shown on the top and bottom rows, respectively, for bimodal networks with $p_{0}=0.95$, $k_{0}=5$ (blue), $K=50$ (red), and $N=500$. Arrows indicate the direction of motion. (a) Translating (b) Ring (c) Hybrid (d) Rotating patterns.}}
\label{fig:Patterns}
\end{figure*}
\begin{figure}[h]
\includegraphics[scale=0.247]{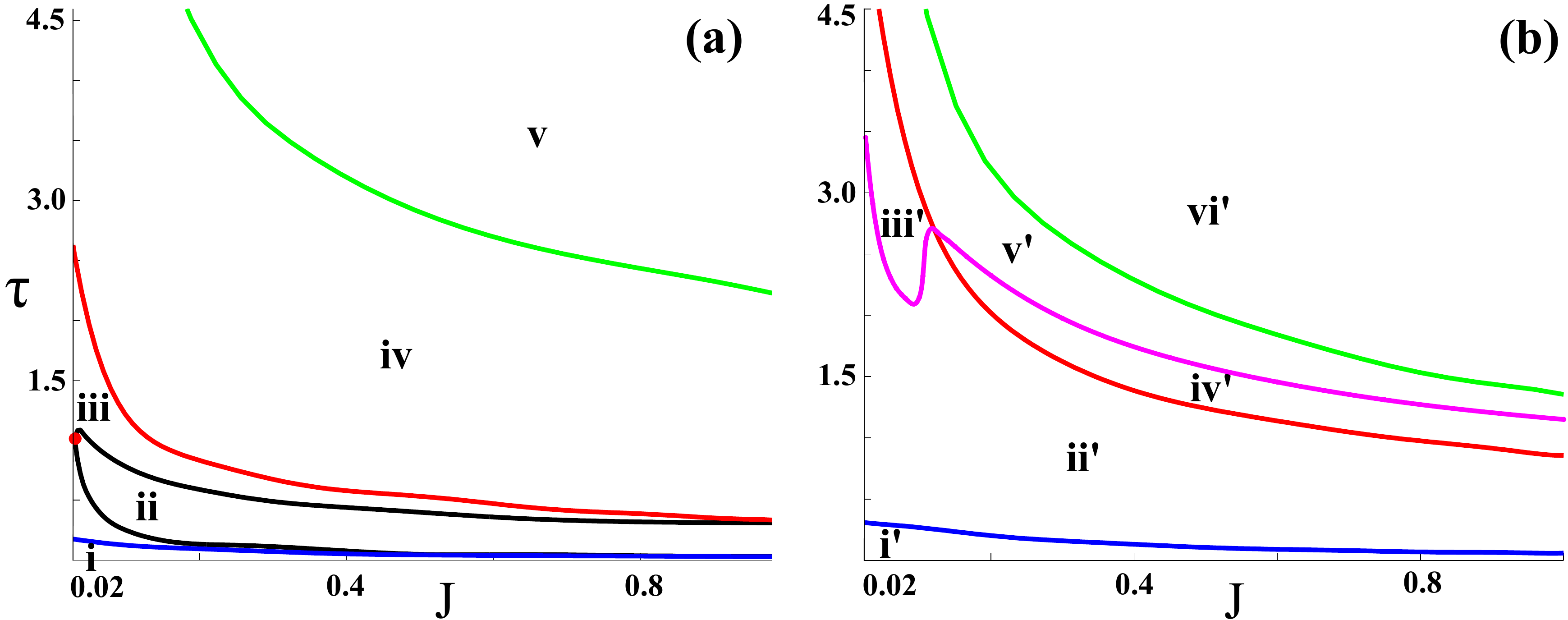}
\caption{{Phase diagram for bimodal networks found by adiabatically changing $J$ and $\tau$ for each pattern until it loses stability\cite{FN3}. (a) heterogeneous network: i. Translating, Ring, and Hybrid states, ii. Ring state, iii. Ring and Hybrid states, iv. Hybrid states, v. Rotating states\cite{FN1}, where $p_{0}=0.9$, $K=60$, $k_{0}=3$ and $N=300$. The red circle marks a $degenerate\!-\!Hopf$ bifurcation. (b) less heterogeneous network: i'. Translating, Ring, and Hybrid states,
ii'. Ring and Hybrid states, iii'. Ring, Hybrid, and Rotating states, iv'. Hybrid states, v'. Hybrid and Rotating states vi'. Rotating states, where $p_{0}=0.95$, $K=25$, $k_{0}=5$, and $N=1000$.}}
\label{fig:PhaseDiagram}
\end{figure}

\section{\label{sec:Annealed} PATTERNS AND DYNAMICS}
In this paper, we study dynamic pattern formation in static networks satisfying predefined degree distributions. We first describe how such a network can be constructed.
Let $p_{k}$ denote the network degree distribution, where the degree, $k$, is the number of links of a node, and $p_{k}$ specifies the fraction of nodes in the network with degree $k$. Networks can be constructed from a prescribed $p_{k}$ with the configuration model ($CM$) by first generating $N$ nodes, each with a number of link ``stubs" drawn from $p_{k}$, and then connecting pairs of ``stubs" to form links, chosen uniformly at random \cite{Newman3}. For simplicity, all links are bidirectional and unweighted, where the connection matrix $A_{ij}=1$ if nodes $i$ and $j$ are linked, and zero otherwise.

Primarily, we focus on bimodal distributions, as a simple construction for networks with both weakly and strongly connected nodes, where $p_{k}$ has a simple form:
\begin{equation}
\label{eq:Distribution}
p_{k}=\left\{
\begin{array}{@{}ll@{}}
    p_{0}, & \text{if}\ k=k_{0} \\
    1- p_{0}, & \text{if}\ k=K \\
    0,  & \text{otherwise,} \\
    \end{array}\right.
\end{equation}
with $k_0 \ll K$. We choose $p_0$ close to $1$ so that agents with large degree $K$ occupy a small portion of the network, and are called ``motherships", while most nodes have degree $k_{0}$ \cite{Hindes, Hindes3}. However, many results are generalized for any $p_{k}$, in which case equations are given in terms of general $k$ and $p_{k}$ (additional example in Sec.\ref{sec:App}).

Given the stated assumptions, a variety of dynamical behaviors are possible depending on the values of coupling strength, $J$, and delay, $\tau$. We first provide
brief descriptions of the basic swarming patterns in Sec.\ref{sec:Description}, and analyze their dynamics in more detail in Sec.\ref{sec:Analysis} with comparisons to simulations.
\subsection{\label{sec:Description} Dynamical behaviors}
Before analyzing the dynamics in detail, it is useful to discuss how different model parameters affect the swarm behavior. In the limit $J\rightarrow0$, all agents are independent and self-propelling, and will travel at unit speed in their initial direction of motion. For relatively small values of $J$ and $\tau$, the propulsion force dominates, and speeds remain near unity. If the swarm has nearly uniform initial conditions, then the coupling will tend to align the directions of motion and favor coherent velocities. This is known as the translating state -- shown in Fig.\ref{fig:Patterns}(a), in which we can see that the entire swarm moves together in the direction of the large arrow, while agents trace out similar orbits (see Sec.\ref{sec:Translation}). On the other hand, if the initial directions are random, then the swarm can organize into a state with no coherent velocity, where propulsion keeps the agents' speeds at unity, but the average directions cancel. This is known as the ring state, which can be seen in Fig.\ref{fig:Patterns}(b). For the bimodal network case shown, we see that agents travel in one of four circular orbits with example directions given by the small arrows.

In general, if $J$ is large such that the spring force is comparable to the self propulsion, then the agents tend to have coherent positions and velocities -- moving together with the swarm's centroid. Moreover, if $\tau$ is also large, then the motion must remain confined -- any large difference between the current and delayed positions would result in a large spring force (Eq.\ref{eq:Dynamics}). This typically leads to coherent rotation, known as the rotating state, which is shown in Fig.\ref{fig:Patterns}(d). Together the three states comprise the known dynamical modes for swarming networks with delay \cite{Romero, Romero2}. Phase diagrams are shown in Fig.\ref{fig:PhaseDiagram} for bimodal networks. Interestingly, several parameter regions contain three stable states. This is a novel feature of swarms with heterogeneous networks.

If the underlying network is very heterogeneous, however, it is possible that different parts of the network may converge to different dynamical modes. For example, for bimodal networks, high and low-degree nodes can split into a state that is a composite of ring and rotating motions -- mixing the behaviors in Fig.\ref{fig:Patterns} (b) and (d), respectively. For example, we find that each degree class's order-parameter (e.g., its centroid) has dynamics analogous to the distinct states \cite{Hindes2}. Therefore, we call this a hybrid state, which is shown in Fig.\ref{fig:Patterns}(c). Detailed descriptions of each state are given in Sec.\ref{sec:Analysis}.

\subsection{\label{sec:Analysis} Analysis}
In order to understand the patterns described in Sec.\ref{sec:Description}, it is useful to treat nodes with the same degree as topologically indistinguishable.
Moreover, we approximate $A_{ij}$ with the weighted average of $CM$s. Since the probability that nodes $i$ and $j$ are linked in a $CM$ is proportional to the product of their degrees, we take $A_{ij}= k_{i}k_{j}/(N\left<k\right>)$. This is known as the annealed network approximation, which allows for qualitatively accurate descriptions of dynamical processes and represents a mean-field theory for heterogeneous networks \cite{Vespignani1}. Analyzing the motion directly from $A$ would result in quantitative improvement, especially in networks with low average degree\cite{FN2}; however, the simple annealed approximation is able to capture much of the behavior.

Let $\mathbf{R}_k$ denote the centroid for each degree class:
\begin{equation}
\label{eq:OP}
\boldsymbol{R}_{k}=\sum_{i | k_{i}=k}\boldsymbol{r}_{i}\Biggr/Np_{k}.
\end{equation}
Given the annealed form for $A$, the equations of motion can be expressed in terms of $\mathbf{R}_k$ as
\begin{align}
\label{eq:DegreeDynamics}
&\ddot{\boldsymbol{r}}_{i}= (1-\mid\!\dot{\boldsymbol{r}}_{i}\!\mid^{2})\dot{\boldsymbol{r}}_{i}-Jk_{i}\Big(\boldsymbol{r}_{i}-\sum_{k}\frac{kp_{k}}{\left<k\right>}\boldsymbol{R}_{k}(t-\tau)\Big),
\end{align}
suggesting Eq.(\ref{eq:OP}) as a useful order-parameter to characterize the net motion of nodes with degree $k$. Comparisons between similar patterns of $CM$ and annealed bimodal networks are shown in the top and bottom rows of Fig.\ref{fig:Patterns}, respectively. The different motions are described in more detail below.

\subsubsection{\label{sec:Ring} Ring state}
\begin{figure}[h]
\includegraphics[scale=0.068]{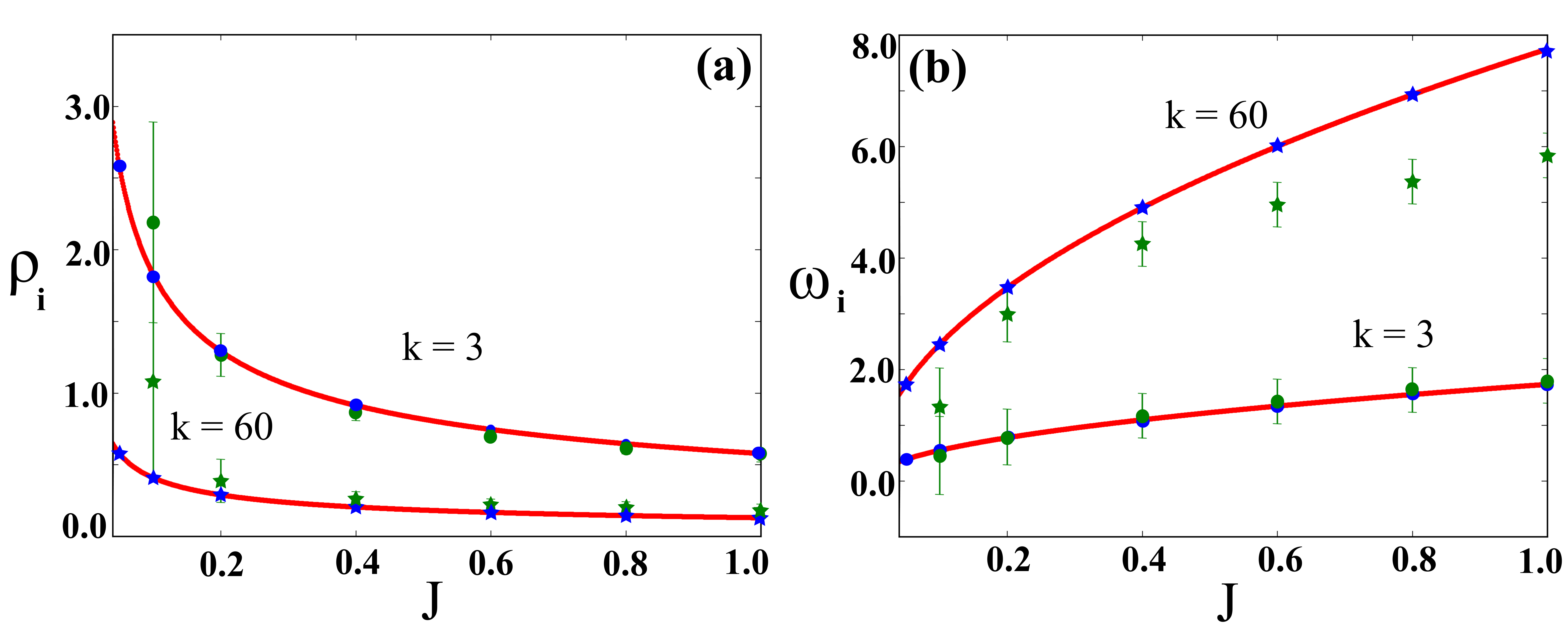}
\caption{{Ring state rotation radii (a) and frequencies (b) for simulated $CM$ (green)
and annealed (blue) bimodal networks compared to predictions (red, Eq.(\ref{eq:Ring})): $\tau\!=\!0.02$, $p_{0}\!=\!0.9$, $k_{0}\!=\!3$, $K\!=\!60$,  and $N\!=\!300$. Parameters correspond to regions beneath the red line in Fig.\ref{fig:PhaseDiagram}(a).}}
\label{fig:Ring}
\end{figure}

For relatively small time delays the ring state is a stable swarm motion pattern.  In the ring state, the agents form concentric rotating rings about a fixed center, such that the swarm has no net motion, $\boldsymbol{R}_{k}\!=\!\boldsymbol{0}$. The radius and angular velocity of the rings depends on the degrees of the constituent agents, as we can find by substituting the ansatz: $\boldsymbol{r}_{i}=(x_{i},y_{i})=\rho_{i}\big[\cos(\omega_{i}t +\phi_{i}),\sin(\omega_{i}t+\phi_{i})\big]$ and $\boldsymbol{R}_{k}\!=\!\boldsymbol{0}$ into Eq.(\ref{eq:DegreeDynamics}):
\begin{align}
\label{eq:Ring}
\rho_{i}=\frac{1}{\sqrt{Jk_{i}}},\qquad \omega_{i}=\pm\sqrt{Jk_{i}}.
\end{align}
This shows that the ring state is composed of pairs of counter-rotating currents for each degree class with unit speed and with radii and frequencies decreasing
and increasing with the square root of the agent degree, respectively (as shown in Fig.\ref{fig:Patterns}(b)). The dependence on degree generalizes homogeneous network results, and in particular, predicts a disordered state with large amplitude and frequency variation for networks with broad $p_{k}$, such as multi-modal or power-law distributions (see Sec \ref{sec:App}) \cite{Klimka}. A comparison between Eq.(\ref{eq:Ring}) predictions and simulation results for bimodal networks are shown in Fig.\ref{fig:Ring} as a function of $J$. Error bars correspond to the standard deviation for each degree class.
\subsubsection{\label{sec:Translation} Translating state}
When the time delay is relatively small, many initial conditions converge to the translating state, in which each degree-class's centroid, Eq.(\ref{eq:OP}), travels at a constant, non-zero velocity. Moreover, for networks with multiple degree classes each centroid is separated in space by some constant displacement from the global center of mass, $\boldsymbol{d_{k}}:$ $\boldsymbol{R}_{k}(t)=\boldsymbol{V}t+\boldsymbol{d_{k}}$, with a velocity $\boldsymbol{V}$. Individual nodes in each degree-class trace out periodic,``bow-tie"-like orbits, as shown in Fig.\ref{fig:Patterns}(a), which is a novel feature of the heterogeneous network pattern.

We can numerically compute the speed and shape of the orbits by inserting the ansatz
$\sum_{k}\frac{kp_{k}}{\left<k\right>}\boldsymbol{R}_{k}(t)=\!\boldsymbol{V}t\;$ into Eq.(\ref{eq:DegreeDynamics}) and
putting all particles in the co-moving frame, $\boldsymbol{z}=\boldsymbol{r}-\boldsymbol{V}t$ (for simplicity, propagation is typically assumed along the line $y=x$, or $\boldsymbol{V}=\big[Vt/\sqrt{2},Vt/\sqrt{2}\;\big])$. This gives a set of single particle ODEs for each degree class, parameterized by the swarm's collective speed:
\begin{align}
\label{eq:Bow-Tie}
&\ddot{\boldsymbol{z}}_{k}= \Big(1-\mid\!\dot{\boldsymbol{z}}_{k}+\boldsymbol{V}\!\mid^{2}\Big)(\dot{\boldsymbol{z}}_{k}+\boldsymbol{V})-Jk\Big(\boldsymbol{V}\tau+\boldsymbol{z}_{k}\Big).
\end{align}

\begin{figure}[t]
\includegraphics[scale=0.34]{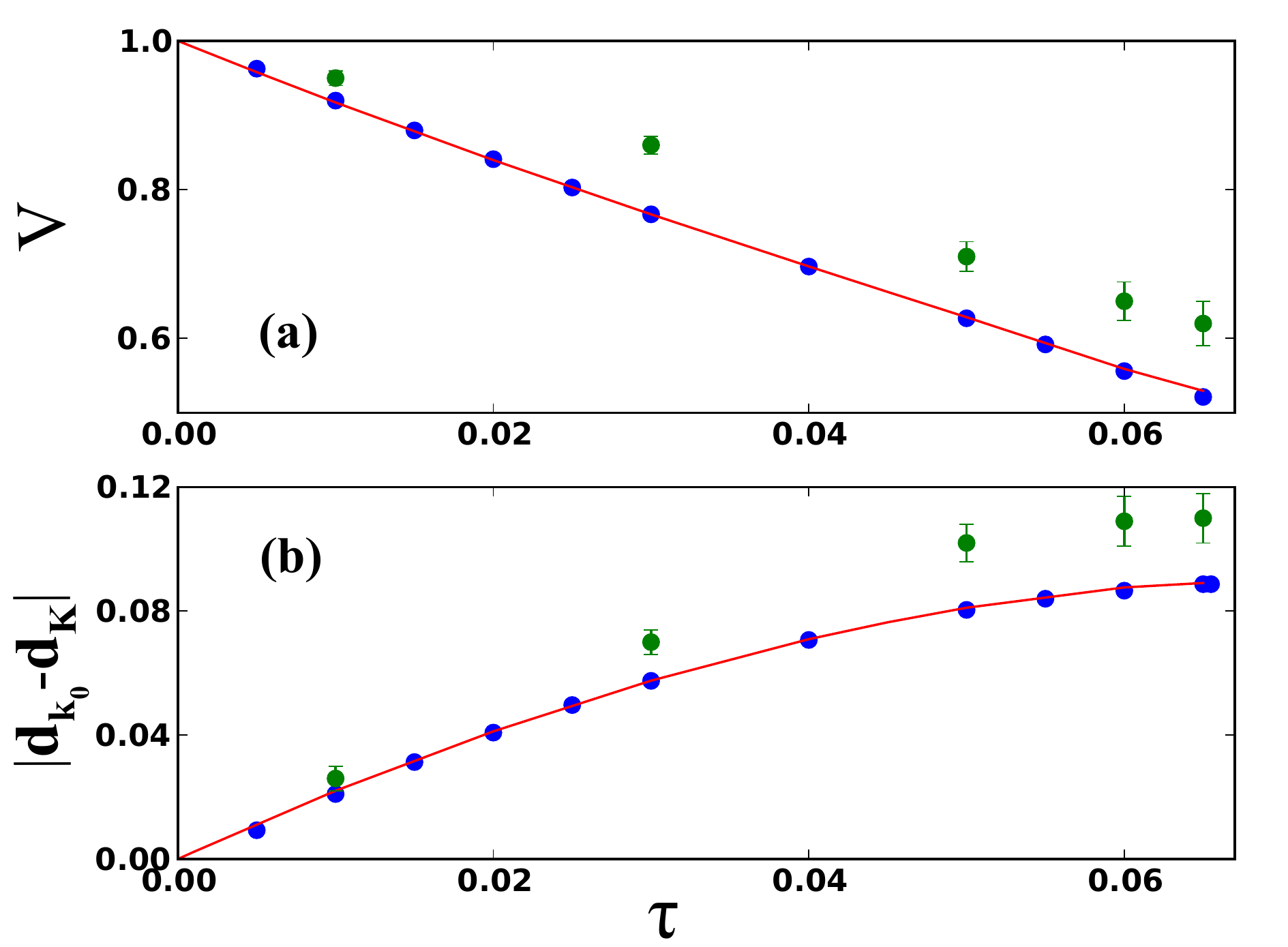}
\caption{{Translating state centroid speeds (a) and displacements (b) for simulated $CM$ (green) and annealed (blue) bimodal networks compared to predictions (red Eq.(\ref{eq:Self-Consistent})) from region $(i)$ in Fig.\ref{fig:PhaseDiagram}(a): $J\!=\!0.333$, $p_{0}\!=\!0.9$, $k_{0}\!=\!3$, $K\!=\!60$, and $N\!=\!300$.}}
\label{fig:Translating}
\end{figure}

In practice, for random initial conditions, Eq.(\ref{eq:Bow-Tie}) has a family of stable ``bow-tie" solutions,
with a $k$-dependent period, $T_{k}$: $\boldsymbol{z}_{k}(t,T_{k};V)$.  These solutions can be used to condition the speed if combined with the self-consistent criterion,
$\sum_{k}\frac{kp_{k}}{\left<k\right>}\boldsymbol{R}_{k}(t)=\!\boldsymbol{V}t\;$ or $\;\sum_{k}\frac{kp_{k}}{\left<k\right>}\boldsymbol{d_{k}}\!=\!\boldsymbol{0}$, by assuming that the swarm density for each degree class is uniform along the orbits, and therefore, replacing $\boldsymbol{d_{k}}$ (the average position from a sum over particles)
with a time average of $\boldsymbol{z}_{k}(t,T_{k};V)$:
\begin{equation}
\boldsymbol{F}(\boldsymbol{V})=\sum_{k}\frac{kp_{k}}{\left<k\right>}\int_{0}^{T_{k}}\frac{\boldsymbol{z}_{k}(t,T_{k};V)dt}{T_{k}}=\boldsymbol{0}.
\label{eq:Self-Consistent}
\end{equation}
For instance, the prediction curve shown in Fig.(\ref{fig:Translating}), was found by generating solutions to Eq.(\ref{eq:Bow-Tie}), $\boldsymbol{z}_{k}(t,T_{k};V)$, from an initial guess for $V$, computing the integral in Eq.\eqref{eq:Self-Consistent}, and updating the guess with a simple Newton method.

Interestingly, we find that the periods are approximately equal to the ring state values, $T_{k}\approx2\pi/\sqrt{Jk}$, as shown in the power spectrum of Fourier modes in Fig.(\ref{fig:Four})(a). This indicates that even though networks have coherent average velocities in the translating state, the individual node dynamics will vary significantly for broad $p_{k}$.
\subsubsection{\label{sec: Rotating} Rotating states}
\begin{figure}[t]
\includegraphics[scale=0.27]{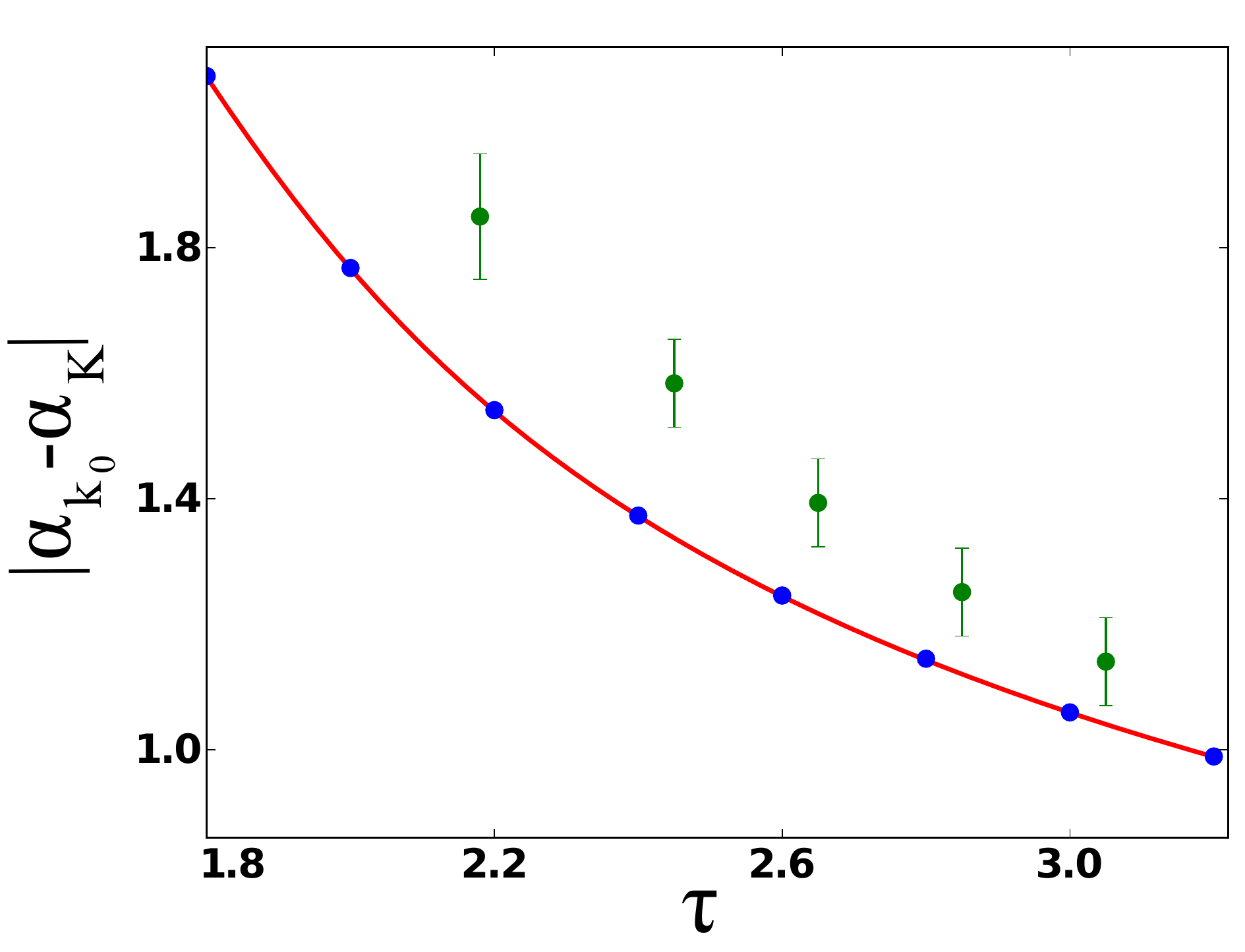}
\caption{{Rotating state phase-differences for simulated bimodal $CM$ (green) and annealed network (blue) swarms compared to predictions (red Eqs.(\ref{eq:Rotation1}-\ref{eq:Rotation2})): $J=0.8$, $p_{0}\!=\!0.95$, $k_{0}\!=\!5$, $K\!=\!25$, and $N\!=\!1000$. Parameters correspond to regions above the magenta line in Fig.\ref{fig:PhaseDiagram}(b).}}
\label{fig:Rotating}
\end{figure}

As explained in Sec.\ref{sec:Description}, for sufficiently large $J$ and $\tau$, all nodes collapse to their respective centroids, such that $\boldsymbol{r}_{i | k_{i}=k} \approx \boldsymbol{R}_{k}$:
\begin{align}
\label{eq:Collapse}
&\ddot{\boldsymbol{R}}_{k}= (1-\mid\!\dot{\boldsymbol{R}}_{k}\!\mid^{2})\dot{\boldsymbol{R}}_{k}\!+\!Jk\Big(\sum_{k'}\frac{k'p_{k'}}{\left<k\right>}\boldsymbol{R}_{k'}(t-\tau)- \boldsymbol{R}_{k}\Big),
\end{align}
with confined rotations about a common center. In general, many dynamical states can satisfy Eq.(\ref{eq:Collapse}). However, simulations from random initial conditions with large $J$ and $\tau$ often converge to a simple frequency synchronized rotation, with amplitudes and phases that vary with degree. Substituting the ansatz $\boldsymbol{R}_{k}(t)\!=\!a_{k}[\cos{\!(\omega_{R} t+\alpha_{k})},\sin{\!(\omega_{R} t+\alpha_{k})}]$, into Eq.(\ref{eq:Collapse}), we find that the synchronized rotation must satisfy:
\begin{align}
\label{eq:Rotation1}
&\!\sum_{k}\!\frac{kp_{k}}{\left<k\right>}a_{k}\!\cos{\!(\!\alpha_{k}\!-\!\omega_{R}\tau\!)} =\\
&\frac{a_{k}}{Jk}\!\Big[\!\big(\!Jk\!-\!\omega_{R}^{2}\big)\!\cos{\alpha_{k}}\!+\!\omega_{R}\big(\!1\!-\!a_{k}^2\omega_{R}^2\big)\!\sin{\alpha_{k}}\!\Big]\!, \nonumber \\
\nonumber \\
\label{eq:Rotation2}
&\!\sum_{k}\!\frac{kp_{k}}{\left<k\right>}a_{k}\!\sin{\!(\!\alpha_{k}\!-\!\omega_{R}\tau\!)} =\\
&\frac{a_{k}}{Jk}\!\Big[\!\big(\!Jk\!-\!\omega_{R}^{2}\big)\!\sin{\alpha_{k}}\!-\!\omega_{R}\big(\!1\!-\!a_{k}^2\omega_{R}^2\big)\!\cos{\alpha_{k}}\!\Big]\!, \nonumber
\end{align}
which generalizes a similar result for the special case of an Erd\H{o}s-R\'{e}nyi network, but for arbitrary $p_{k}$ (see Fig.\ref{fig:Rotating}), and predicts a broad range of amplitudes and phases for very heterogeneous networks, such as multi-modal or power-law $p_{k}$ (see Sec \ref{sec:App}) \cite{Klimka}. In general, Eqs.(\ref{eq:Rotation1}-\ref{eq:Rotation2}) must be solved numerically and have many solutions depending on the parameters, though most are found to be unstable.

Additionally, we find that such frequency synchronized rotations emerge through a set of $Hopf$ bifurcations of $\bold{r}_{j}\!=\!\bold{0}$,  where perturbations with uniform amplitudes and k-dependent phases, $\boldsymbol{r}_{j}\!=\!\boldsymbol{\epsilon}e^{i(\alpha_{k_{j}}+\omega_{R}t)}$, are dynamically neutral to linear order in $\epsilon$ with $\omega_{R}\neq0$. The general $p_{k}$-dependent form of the $Hopf$ bifurcation for synchronized rotations can be found by taking $a_{k}\!\rightarrow \!a\! \rightarrow \!0$ in Eqs.(\ref{eq:Rotation1}-\ref{eq:Rotation2}), solving for $\cos{\alpha_{k}}$ and $\sin{\alpha_{k}}$, multiplying by $kp_{k}/\!\left<k\right>$, summing over $k$, and eliminating the $k$-independent constants $\sum_{k}\!\frac{kp_{k}}{\left<k\right>}\!\cos{\alpha_{k}}$ and $\sum_{k}\!\frac{kp_{k}}{\left<k\right>}\!\sin{\alpha_{k}}$, giving:
\begin{align}
\label{eq:RotEmer1}
\tan(\omega_{R}\tau)&= \frac{\omega_{R}}{J\left<k\right>-\omega_{R}^{2}}, \\
\label{eq:RotEmer2}
\!\Bigg|\!\sum_{k}\!\frac{kp_{k}}{\left<k\right>}\!e^{i\alpha_{k}}\Bigg|^{2}\!&= \sqrt{\!\Bigg(1-\frac{\omega_{R}^{2}}{J\left<k\right>}\Bigg)^{2}\!\!\!+\!\Bigg(\frac{\omega_{R}}{J\left<k\right>}\Bigg)^{2}}.
\end{align}
%Examples are shown in Fig.\ref{fig:PhaseDiagram} \textcolor{blue}{(color?)} for the simple case of a bimodal network.
In general, Eqs.(\ref{eq:RotEmer1}-\ref{eq:RotEmer2}) can specify existence conditions for synchronized rotations, but not necessarily stability, and therefore only bound the region above the magenta line in Fig.\ref{fig:PhaseDiagram}(b), for example.

\subsubsection{\label{sec:Hybrid}Hybrid states}
\begin{figure}[h]
\includegraphics[scale=0.068]{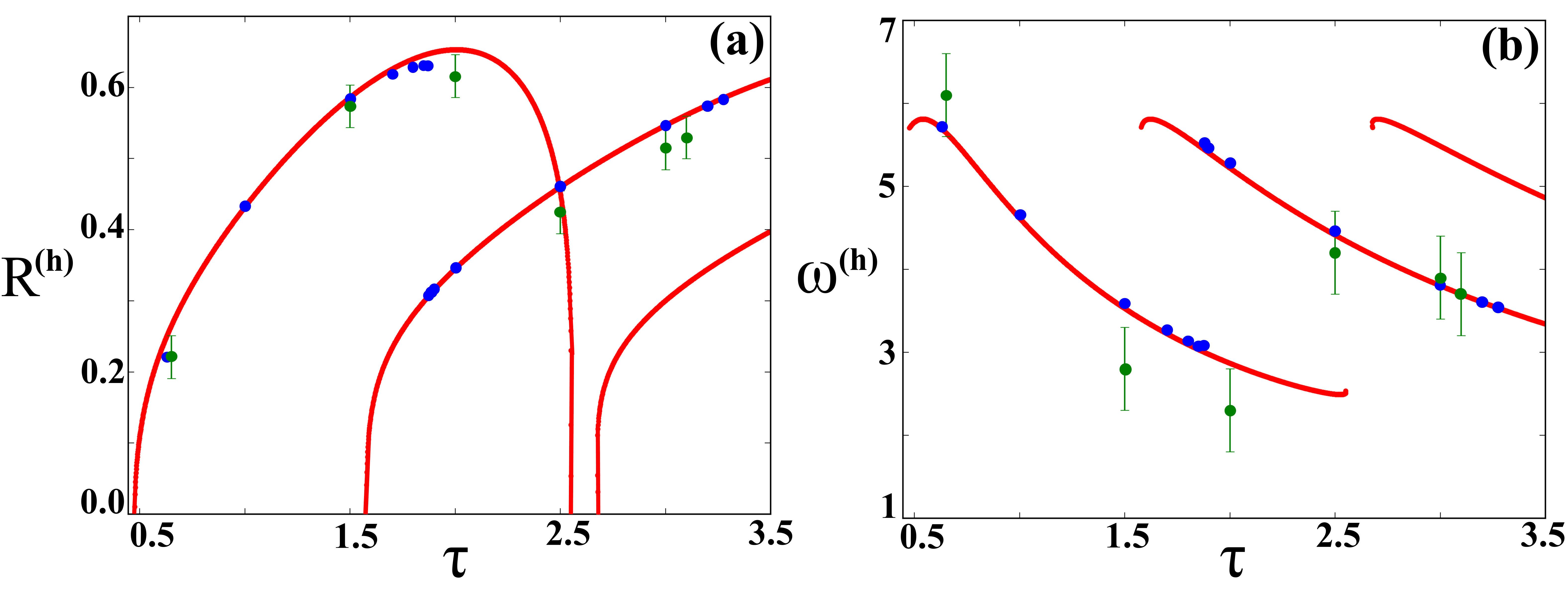}
\caption{{Mothership rotation radius (a) and frequency (b) in the hybrid state for simulated $CM$ (green)
and annealed (blue) bimodal networks compared to predictions (red, Eqs.(\ref{eq:Hybrid}-\ref{eq:Hybrid2})): $J\!=\!0.333$, $p_{0}\!=\!0.9$, $k_{0}\!=\!3$, $K\!=\!60$,  and $N\!=\!300$. Parameters correspond to regions (iii) and (iv) in Fig.\ref{fig:PhaseDiagram}(a).}}
\label{fig:Hyb}
\end{figure}

As hinted in Sec.\ref{sec:Description} and shown in Fig.\ref{fig:PhaseDiagram}, for both large and small delays hybrid motions can be stable, in which high-degree nodes collapse to their centroid and rotate approximately uniformly with a constant radius and frequency, while weakly driving low-degree nodes around a motion that is similar to the ring state. By neglecting the small coherence from low-degree nodes and looking for solutions of Eq.(\ref{eq:DegreeDynamics}): $\boldsymbol{R}_{k_{0}}\!=\!\boldsymbol{0}$ and $\boldsymbol{r}_{i | k_{i}=K}\!=\!\boldsymbol{R}_{K}(t)\!=\!R^{(h)}(\cos(\omega^{(h)}t),\sin(\omega^{(h)}t)$, we find the hybrid rotation satisfies
\begin{align}
\label{eq:Hybrid}
&{\omega^{(h)}}^{2} = JK\Bigg(1-\frac{K(1-p_{0})}{\left<k\right>}\cos{\omega^{(h)}\tau}\Bigg), \\
\label{eq:Hybrid2}
&\omega^{(h)}(1-{R^{(h)}}^{2}{\omega^{(h)}}^{2})\!=\!  \frac{JK^{2}(1\!-p_{0})}{\left<k\right>}\sin{\omega^{(h)}\tau},
\end{align}
where the two centroids have dynamics analogous to the ring and rotating states simultaneously. Like the rotating state, many solutions are possible to Eqs.(\ref{eq:Hybrid}-\ref{eq:Hybrid2}) in general, depending on the parameter values, including multiple stable branches. This can lead to discontinuous jumps between hybrid states with different frequencies (as shown in Fig.\ref{fig:Hyb}).

\begin{figure}[t]
\includegraphics[scale=0.245]{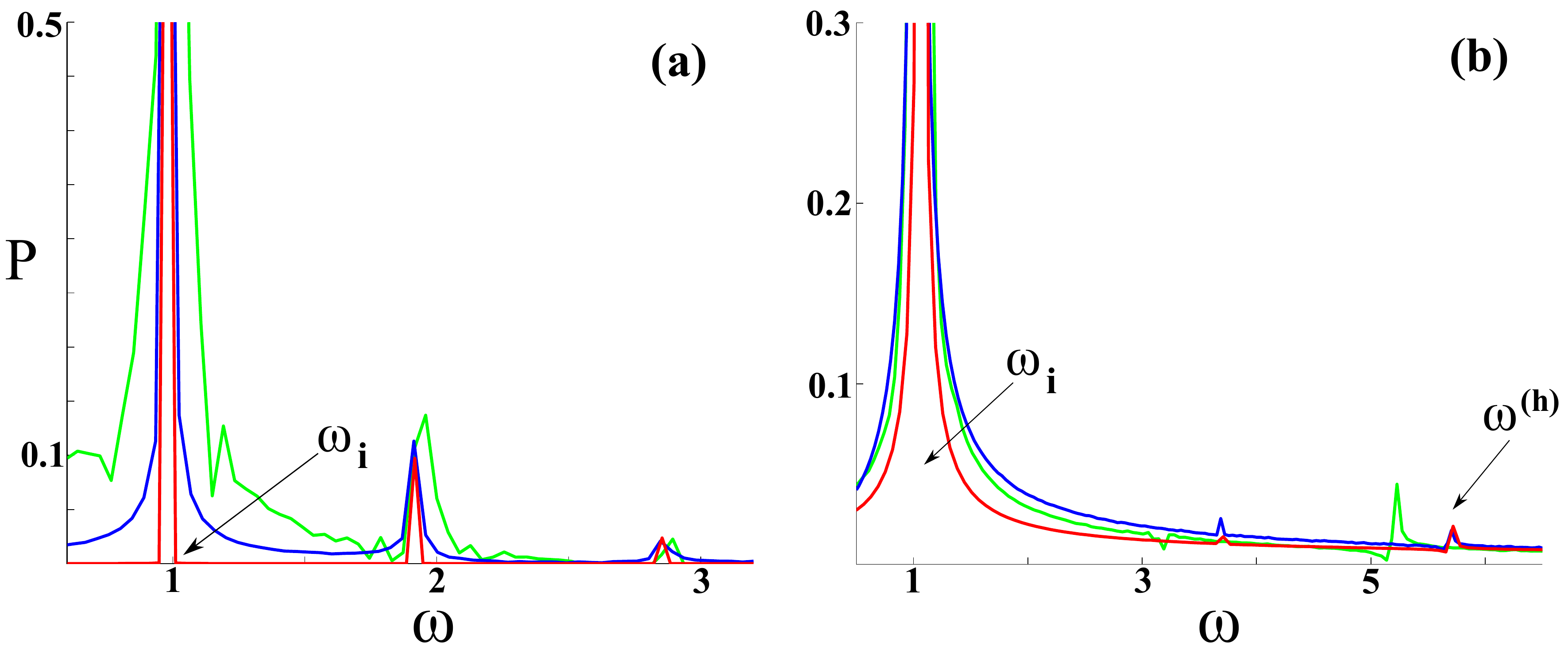}
\caption{Fourier$(\omega)$ power spectra of low-degree nodes for simulated $CM$ (green) and annealed (blue) bimodal networks. (a) translating state
spectrum for $\tau\!=0.05$ compared to predictions (red (Eq.\ref{eq:Self-Consistent})). (b) hybrid state spectrum for $\tau\!=0.65$ compared to predictions (red (Eq.\ref{eq:HybridLow})). Large peaks in (a) and (b) correspond to the ring frequency. $J\!=\!0.333$, $p_{0}\!=\!0.9$, $k_{0}\!=\!3$, $K\!=\!60$, and $N\!=\!300$.}
\label{fig:Four}
\end{figure}

On the other hand, the low-degree node dynamics can be found by substituting the mothership rotation from Eqs.(\ref{eq:Hybrid}-\ref{eq:Hybrid2}) into Eq.(\ref{eq:DegreeDynamics}). This gives a four dimensional set of single-particle ODEs to be integrated:
\begin{align}
\label{eq:HybridLow}
&\ddot{\boldsymbol{r}}-(1-\mid\!\dot{\boldsymbol{r}}\!\mid^{2})\dot{\boldsymbol{r}}+ Jk_{0}\boldsymbol{r}= Jk_{0}\frac{K(1-p_{0})}{\left<k\right>}\boldsymbol{R}^{(h)}(t-\tau).
\end{align}
The expected form of the dynamics -- a ring-like motion driven by a periodic force -- is found by examining the left and right hand sides of Eq.(\ref{eq:HybridLow}). In particular, when $\boldsymbol{R}_{K}\rightarrow\boldsymbol{0}$, the equations of motion for a ring state are recovered.  Both dynamical signatures can be seen clearly in the power spectrum of Fourier modes of Eq.(\ref{eq:HybridLow}), which has a large peak at the ring frequency, $\omega_{i}$, and a small peak at the hybrid frequency, $\omega^{(h)}$. Comparisons between the predicted and simulated dynamics for the hybrid state are shown in Fig.(\ref{fig:Four})(b) and Fig.(\ref{fig:Hyb}). In general, two rotation directions are possible simultaneously depending on the initial conditions for Eq.(\ref{eq:HybridLow})-- similar to the ring state.

In addition, we can find approximately where hybrid states emerge, and thus bound their stability regions in Fig.\ref{fig:PhaseDiagram}, by taking $R^{(h)}\rightarrow0$ in Eq.(\ref{eq:Hybrid2}). This is coincident with another set of $Hopf$ bifurcations of $\bold{r}_{j}\!=\!\bold{0}$ (in addition to those corresponding to rotating states), where perturbations $\boldsymbol{r}_{j}\!=\!\boldsymbol{\epsilon}e^{(i\omega^{(h)}t)} \delta_{k_{j},K}$ are dynamically neutral to linear order in $\epsilon$ with $\omega^{(h)}\neq0$. Eliminating $\tau$ in Eqs.(\ref{eq:Hybrid}-\ref{eq:Hybrid2}) gives a polynomial expression for the bifurcation frequency, $\omega_{*}^{(h)}$:
\begin{align}
\label{eq:HybridEmergence}
\Bigg(1-\frac{{{\omega_{*}^{(h)}}}^{2}}{JK}\Bigg)^{2}+\Bigg(\frac{\omega_{*}^{(h)}}{JK}\Bigg)^{2}= \Bigg(\frac{K(1-p_{0})}{\left<k\right>}\Bigg)^{2},
\end{align}
that can be  combined with Eq.(\ref{eq:Hybrid}) to predict the black curves in Fig.\ref{fig:PhaseDiagram}(a).

Interestingly, Eq.(\ref{eq:HybridEmergence}) has degenerate solutions for ${\omega_{*}^{(h)}}$, if
\begin{align}
\label{eq:Degeneracy}
 1=4JK\Bigg[1-JK\bigg(\frac{K(1-p_{0})}{\left<k\right>}\bigg)^{2}\Bigg],
\end{align}
corresponding to $degenerate\!-\!Hopf$ bifurcations, shown in Fig.(\ref{fig:PhaseDiagram}) where the $Hopf$ bifurcations meet.\\

%in addition to those corresponding to hybrid states. However, in this case the bifurcations occur

\section{\label{sec:Disc}DISCUSSION}
In this work we studied swarming of self-propelled autonomous agents with time-delayed interactions over
heterogeneous networks. For many swarm models in biology, emergent behavior due
to coupling includes a basis of dynamical patterns, such as translation and
ring dynamics about a stationary center of mass \cite{Erdmann}. In addition, if the coupling
communication between agents is delayed, a rotating state appears in which
the agents are highly aligned and localized \cite{Romero2,Forgoston}.

The current research builds on the previous homogeneous network swarms, by
generalizing the network topology. In particular, in contrast to all-to-all
coupling, we consider communication networks with finite degree chosen from
a given distribution.  One interesting distribution we considered was
bimodal, in which the network was constructed with a few high-degree nodes and a large number of low-degree nodes.
The topology is a cross between a star network in which all agents
communicate through a single ``mothership", and all-to-all
communication with no special nodes. For the bimodal topology, we described novel
hybrid patterns, both numerically and analytically, consisting of a nonlinear combination of
basis modes from homogeneous networks. In particular, we found a state where high and low-degree nodes have simultaneous dynamics
that are analogous to the rotating and ring states, respectively.
Though relatively simple in this case, we suggest that hybrid behaviors may be a general feature of nonlinear
processes on networks with highly heterogeneous communication topologies, where the local order-parameters for
parts of a network have qualitative differences in their dynamics, corresponding to separate states in homogeneous networks.

In addition, we demonstrated how to generalize several known patterns for
networks with general degree distributions\k{,} including the translating, ring,
and rotating states. This was done by applying a mean-field approximation
scheme, which enabled us to develop lower-dimensional analytic and numerical
procedures for computing aspects of the swarming patterns, such as the
amplitudes, phases, and frequencies of rotation, and the velocities of
translating states. Similar techniques may be generally useful for other
nonlinear problems on networks with delay. Predictions were compared to both
quenched and annealed network simulations with good agreement. We note that in addition to those states predicted here, there exist
several other complex states which appear as a result of the infinite dimensional
dynamics of the delay coupled network. The full unfolding of these states is beyond the scope of this
work, but is of interest when considering basins of attraction of the states discussed.

Since we can port our model to a real experimental workspace, as a next step we plan to realize the predicted patterns
in both two-wheeled and quad-rotor robotic swarms \cite{Klimka}. Further
experiments will lead to interesting questions, such as how to design
parametric controls that can steer a swarm among targeted behaviors in real
environments by exploiting topology. Since fluctuations and uncertainty are an inevitable feature of most
environments, it will be necessary to understand the effects of noise on
swarming dynamics, and how different networks respond to
fluctuations \cite{Erdmann, Forgoston, Lindley3}. Controlling networks with stochastic dynamics is a rich area for
practical and theoretical research \cite{Hindes3, Motter}.
%In addition, we plan to study swarming patterns and stability in non-random
%network structures, such as self-similar topologies where the system is composed of nested layers of swarming ``teams."
%\ira{IBS: Use the in the above paragraph on generalizing the theory.}
%In addition, we plan to study swarming patterns and stability in more general
%network structures. Numerical evidence suggests that the techniques developed
%here pertain, as well, to more general heterogeneous networks, such as
%power-law degree distributed. Studying swarming in non-random, self-similar
%topologies ( e.g., hierarchies) is also of interest, in which case the
%communication network is composed of nested layers of swarming ``teams." Both
%topological features may be useful for developing robotic swarms to perform
%tasks over many space and time scales simultaneously.

\section{\label{sec:Ack}ACKNOWLEDGMENTS}
\noindent J. H. and K. S. are National Research Council postdoctoral fellows. I.B.S was supported by the U.S. Naval Research Laboratory funding (N0001414WX00023) and office of Naval Research (N0001416WX00657) and (N0001416WX01643).

\section{\label{sec:App}APPENDIX}
\begin{figure}[h]
\includegraphics[scale=0.2315]{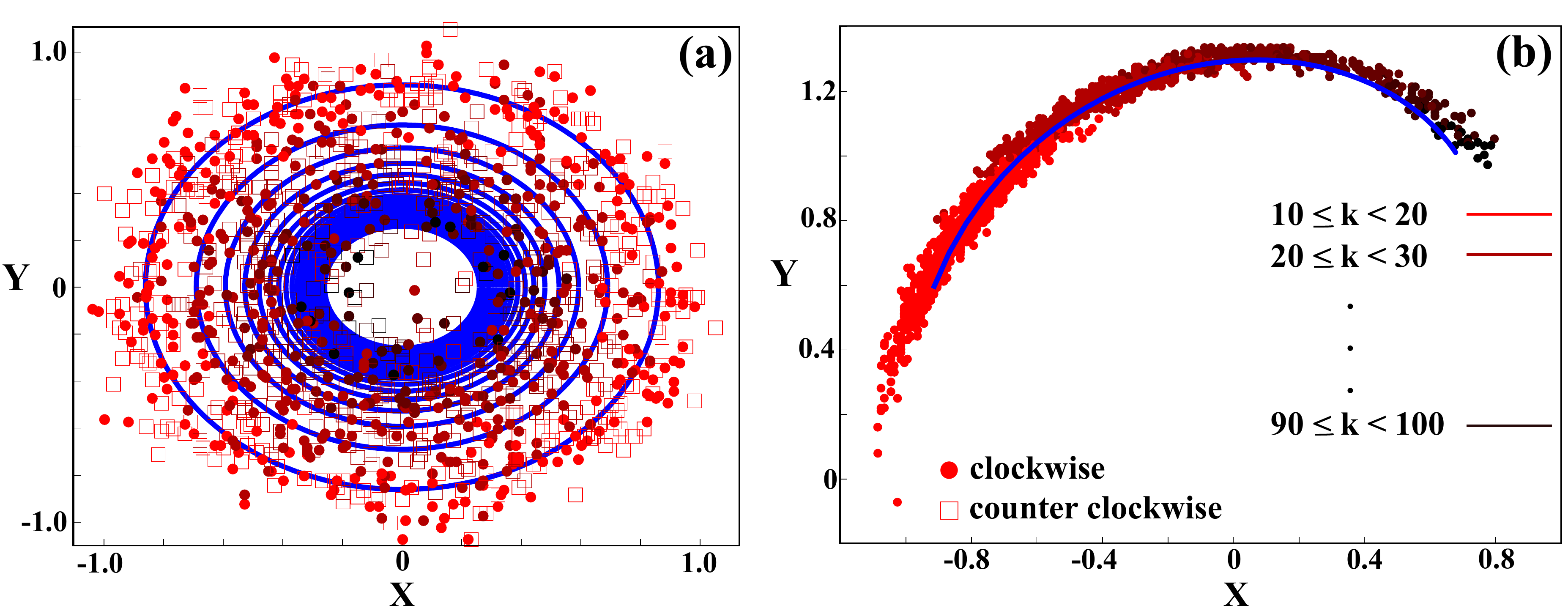}
\caption{{Spatial patterns for a power-law network in the ring (a) and rotating (b) states. (a) Ring state where clockwise and counter clockwise rotation are shown with red circles and squares, respectively, and with colors darkening with increasing $k$ (legend in (b)). Predictions from Eq.(\ref{eq:Ring}) are shown in blue for degree classes in multiples of $5$, i.e., $k=10,15...100$. Simulation parameters are $N=1000$, $J=0.15$, and $\tau=1.5$. (b) Rotating state with the same legend as (a). Prediction from Eqs.(\ref{eq:Rotation1}-\ref{eq:Rotation2}) are shown in blue for all $k$. Simulation parameters are $N=1000$, $J=0.15$, and $\tau=4.0$.}}
\label{fig:RotPL}
\end{figure}

The above comparisons between mean-field predictions and network simulations were restricted to bimodal networks for clarity, though many results were stated for general distributions. As an example, we show a $CM$ network with a truncated power-law degree distribution, $p_{k}=\!k^{-2.5}\!/\!\sum_{k'=10}^{100}k'^{-2.5}$,  in the ring and rotating states in Fig.\ref{fig:RotPL}.  A similar comparison can be done for the translating state, and there is some numerical evidence for the existence of hybrid motion, but a more complete analysis remains for future work.

\begin{thebibliography}{9}
\bibitem{FN1} The rotation in Fig.\ref{fig:PhaseDiagram}(a) satisfies Eq.(\ref{eq:Collapse}) but is not a simple circular motion as described in Sec.\ref{sec: Rotating}.
\bibitem{FN2} Spectral techniques that employ $A$ directly can provide quantitative improvements in accuracy, e.g, in networks with large spectral gaps.
\end{thebibliography}
\end{document}